\documentclass[11pt,onecolumn]{article}
\usepackage{amsmath}
\usepackage{amsfonts}
\usepackage{amssymb}

\usepackage[latin1]{inputenc}
\usepackage{latexsym} 
\usepackage{graphicx}
\usepackage{psfrag}
\usepackage{exscale}
\usepackage[permil]{overpic}       
\usepackage[vflt]{floatflt}        
\usepackage{times}
\usepackage{makeidx}               
\usepackage{color}
\usepackage{framed,color}
\usepackage{hyperref}

\makeindex   

\newfont{\Head}{cmss17 at 36pt}
\newfont{\head}{cmss17}
\newfont{\auflage}{cmss10 at 14.4pt}
\input epsf 
\epsfverbosetrue 

   {\begin{list}{}{
      \settowidth{\labelwidth}{\textbf{#1}}
      \setlength{\leftmargin}{\labelwidth}
         \addtolength{\leftmargin}{\labelsep}
      \setlength{\parsep}{0.5ex plus0.2ex minus0.2ex}
      \setlength{\itemsep}{0.3ex}
      }}
   {\end{list}}

\newcommand{\ze}[2]{#1\;{\rm#2}} 

\newcommand{\zze}[3]{#1\cdot 10^{#2}\;{\rm#3}} 

\newfont{\tensy}{cmsy10}           

\newcommand{\grad}{^\circ} 
                                        
\newcommand{\fm}[1] 
{\begin{displaymath}#1\end{displaymath}}

\newcommand{\nfm}[1] 
{\begin{equation}#1\end{equation}}

\newcommand{\nmzfm}[1] 
{\begin{eqnarray}#1\end{eqnarray}}

\newcommand{\mzfm}[1] 
{\begin{eqnarray*}#1\end{eqnarray*}}

\newcommand{\gfm}[1] 
{\begin{displaymath}\fbox{$\displaystyle #1$}\end{displaymath}}

\newcommand{\ngfm}[1] 
{\begin{equation}\fbox{$\displaystyle #1$}\end{equation}}

\newcommand{\mzgfm}[1] 
{\begin{displaymath}\fbox{$\begin{array}{rcl} #1\end{array}$}\end{displaymath}}

\newcommand{\figcapt}[2] 
{\renewcommand{\normalsize}{\small}\textbf{\caption[#1]{\textmd{#2}}}}

\newcommand{\fif}[1] 
{\mathbf{#1}}

\newcommand{\bi}[4] 
{\int_{#1}^{#2}{#3}\,{#4}}

\newcommand{\ubi}[2] 
{\int{#1}\,{#2}}

\definecolor{shadecolor}{gray}{.90} 

\newtheorem{beispiels}{Beispiel}[section] 

\newtheorem{aufgabeS}{Aufgabe}[section]

\newcommand{\qM} 
{\left[\ddots\right]}

\newcommand{\SV} 
{\left[\vdots\right]}

\title{Open Photoacoustic Cell for Blood Sugar Measurement: Numerical Calculation of Frequency Response}

\author{\normalsize Bernd Baumann\thanks{E-mail: info@BerndBaumann.de} \and \normalsize Marcus Wolff \and \normalsize Mark Teschner}

\date{\small  Hamburg University of Applied Sciences, Heinrich Blasius Institute of Physical Technologies, Department of Mechanical Engineering and Production, \newline  Berliner Tor 21, 20099 Hamburg, Germany}

\begin{document}

\maketitle

\vspace{3mm}
{\noindent\bf Abstract:} 
A new approach for continuous and non-invasive monitoring of the glucose concentration in human epidermis has been suggested recently. This method is based on photoacoustic (PA) analysis of human interstitial fluid. The measurement can be performed in vitro and in vivo and, therefore, may form the basis for a non-invasive monitoring of the blood sugar level for diabetes patients. It requires a windowless PA cell with an additional opening that is pressed onto the human skin. Since signals are weak, advantage is taken of acoustic resonances of the cell. Recently, a numerical approach based on the Finite Element (FE) Method has been successfully used for the calculation of the frequency response function of closed PA cells. This method has now been adapted to obtain the frequency response of the open cell. Despite the fact that loss due to sound radiation at the opening is not included, fairly good accordance with measurement is achieved.

\vspace{3mm}
{\noindent\bf Keywords:} Photoacoustics; acoustic resonator; acoustic loss; blood sugar concentrations

\newpage
\section{Introduction}\label{Introduction}

The photoacoustic or photothermal effect is utilized for the highly sensitive detection of weakly absorbing samples \cite{Haisch2012,BozokiEtAl2012,BageshwarEtAl2010,Michaelian2003}. In photoacoustic sensors (PAS) radiation from a source, typically an infrared laser, is used to excite vibrational states of molecules. Non-radiating relaxation of these molecules leads to a local elevation of the temperature. Since the radiation is modulated, the temperature varies periodically. The temperature variation is accompanied by a pressure modulation, which can be detected by a microphone. As long as no saturation occurs, the PA signal is proportional to the concentration of the excited molecules.

To obtain a large PA signal one often uses the closed sample cell as an acoustic resonator. When the modulation frequency is tuned to an eigenfrequency of this resonator, acoustic modes are excited which results in signal amplification.

Recently it has been suggested to use PAS for the non-invasive in vivo measurement of the glucose content in the interstitial fluid. The glucose concentration in the interstitial fluid is correlated to the glucose level of the blood and the goal is to obtain a fast, cheap and painless measurement of the blood sugar level of diabetics \cite{PleitezEtAl2013, Kottmann2012}. This would also be a large step forward to a sufficient monitoring of blood sugar content as required by physicians.

\begin{figure*}
\centering
{\includegraphics[width=0.45\textwidth,trim=0 0 0 50,clip]{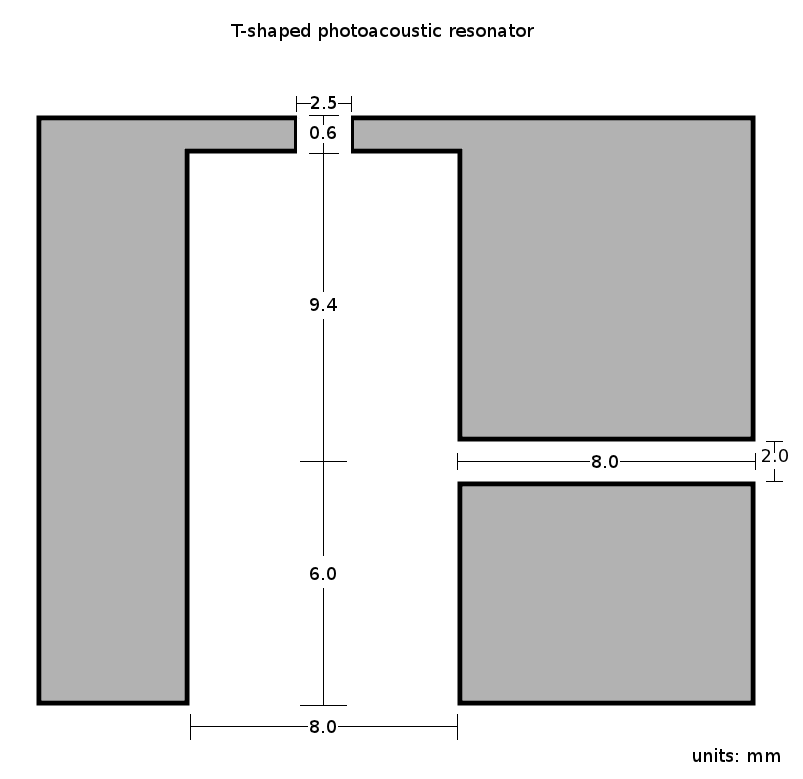}}{\includegraphics[width=0.45\linewidth,trim=25 30 50 20 ,clip]{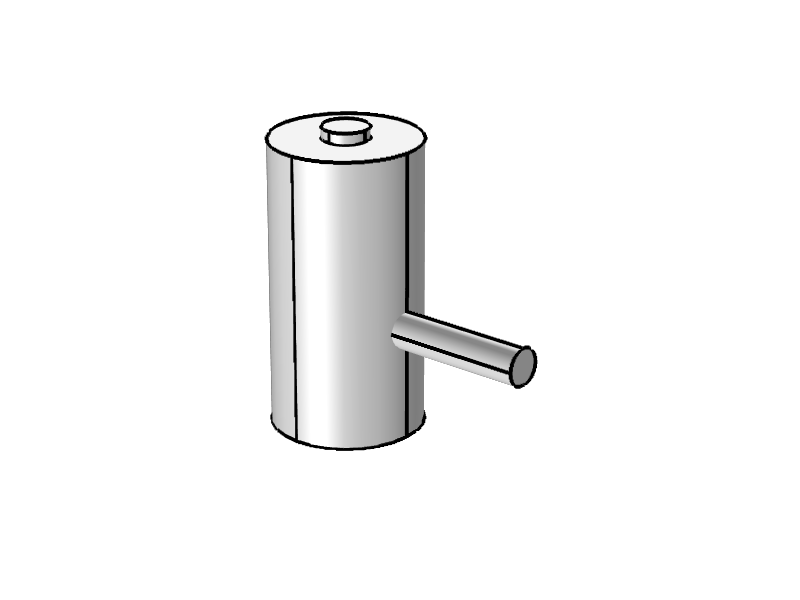}}
\caption{T shaped resonator of the PA sensor \textcolor{black}{made from stainless steel}. In the first experiments the opening at the bottom of the large cylinder (main cylinder) was closed with an optical window (closed resonator). The small opening at the top of the cell is closed by the skin of the patient. The microphone is mounted at the end of the narrow cylinder (resonance cylinder).}
\label{fig:FrankfurtResonator}
\end{figure*}

The T shaped resonator of the PAS for blood sugar monitoring introduced in \cite{PleitezEtAl2013} is depicted in Figure \ref{fig:FrankfurtResonator}. It consists of a broad main cylinder with a base which is either closed by a window or open (laser beam opening). The axis of the narrow resonance cylinder is perpendicular to the axis of the main cylinder. The microphone is mounted near it's outer end. The probands have to close the absorption opening at the top of the main cylinder by pressing the skin of their hand to the cell. After entering the cell through the optical window respectively through the laser beam opening the modulated laser radiation (\"Uber Tuner 9, Daylight Solutions, California) traverses the PA cell and is \textcolor{black}{directed onto the skin surface}. The interstitial fluid which contains the glucose molecules is located about $\ze{50}{\mu m}$ to $\ze{100}{\mu m}$ below the surface. From here the PA sound waves are emitted, travel into the resonator and are detected by the microphone.

Early tests in ambient air with such a device revealed a problem: Due to skin transpiration during the measurements humidity accumulated inside the temperature stabilized cell leading to condensation at the cell walls and the  window. However, humidity in the sample cell  increases the attenuation of sound waves especially at frequencies above $\ze{10}{kHz}$ \cite{MiklosLoerincz1989}. One remedy could be to guide a gentle flow of dry nitrogen through the cell \cite{Kottmann2012}. Unfortunately, this flow results in noise, which superimposes the PA signal. \textcolor{black}{ In addition, the necessity to amend the device by a fixture for the maintenance of a steady nitrogen flow is in conflict to the aim of a simple blood sugar sensor for domestic us. An alternative solution uses an open resonator and thus allows the water molecules to evaporate out of the cell. An open resonator has also the advantage of keeping the static pressure at a constant value and temperature variations due to skin-cell contact small. On the other hand, the opening of the resonator deteriorates the signal and, therefore, the detection sensitivity.}

However, a strong PA signal is of crucial importance. The determining factor for the signal strength is the shape of the resonator. Unfortunately, it is not obvious which cell geometry results in a high acoustic amplification. Testing different shapes experimentally is an option but  would require much time and is expensive. It is considerably more efficient to calculate the signal strength of a certain geometry. A method for the analytical calculation of the PA signal has been available for a long time \cite{Kreuzer1977}. The method, however, is applicable for  simple geometries like cylinders only. In \cite{BaumannEtAl2007} we have combined the method of Kreuzer with the Finite Element Method (FEM), thus allowing to apply it to virtually arbitrary shapes of resonators. The procedure used to be restricted to closed resonators. Here, we investigate the possibility to extend  the application to open resonators.


\section{Theoretical background}\label{Sec:TheoreticalBackgroung}

\subsection{Calculating the PA signal for closed  resonators}\label{Subsec:ClosedResonator}

The acoustic pressure $p$ at the microphone position\footnote{In principal, one needs to integrate the acoustic pressure over the diaphragm of the microphone. Nowadays diaphragms are very small and it is sufficient to calculate the pressure at one representative point.}$\vec{r}$ and a modulation frequency $\omega$ can be determined by solving the inhomogeneous Helmholtz equation
\nfm{\vec{\nabla}^2 p(\vec{r},\omega)+ k^2 p(\vec{r},\omega)={\rm i}\omega\frac{\gamma-1}{c^2}{\cal H}(\vec{r},\omega) \,.\label{dgl1}}
$c$ is the speed of sound, $k$ the acoustic wave number and ${\cal H}(\vec{r},\omega)$ the Fourier transform of the power density $H(\vec{r},t)$. $\gamma$ is the ratio of the isobaric and isochoric heat capacities. If the absorption of the radiation by the molecules is not saturated and the modulation frequency is much smaller than the relaxation rate of the molecular transition ${\cal H}(\vec{r},\omega)=\alpha {\cal I}(\vec{r},\omega)$, where ${\cal I}(\vec{r},\omega)$ is the Fourier transformed intensity of the electromagnetic field.  $\alpha$ is the absorption coefficient for the infrared radiation in the sample. The walls of the resonator are assumed to be sound hard. 
The solution of Equation (\ref{dgl1}) can be expressed as
\nfm{p(\vec{r},\omega)=\sum_{j}^{}{A_j(\omega)p_j(\vec{r}),}\label{eq:solution1}}
where the modes $p_j(\vec{r})$ and the according eigenfrequencies $\omega_j=ck_j$ are obtained by solving the homogeneous Helmholtz equation. The modes are orthogonal and have to be normalized. 

The frequency dependency of the amplitudes $A_j(\omega)$ in Equation (\ref{eq:solution1}) has the form
\nfm{A_j(\omega)=
{\rm i}\frac{{\cal A}_j\omega}{\omega^2-\omega_j^2}.\label{eq:Amplitudes1}}
The contribution of a certain mode is determined by the excitation amplitude 
\nfm{{\cal A}_j=
\frac{\alpha(\gamma-1)}{V_{\rm C}} \int_{V_{\rm C}}{}{p_j^\ast\cdot {\cal I}}{{\;\rm d}V},\label{eq:Amplitudes2}}
where $V_{\rm C}$  is the volume of the PA cell and $p_i^\ast$ the conjugate-complex of $p_i$.
As mentioned in the introduction the sound is excited a small distance below the skin surface. In principle, the domain of integration has to be extended to the region, where the radiation is absorbed. In practice, this has been handled a little different (see Section \ref{Subsec:ResultsClosedResonator}).

So far the model described above does not account for loss. If loss is small it is possible to incorporate the damping of sound waves through the inclusion of loss factors $\ell_j$ in the amplitude Equation (\ref{eq:Amplitudes1}):
\nfm{A_j(\omega)\rightarrow A_j(\omega)=
{\rm i}\frac{{\cal A}_j\omega}{\omega^2-\omega_j^2+{\rm i}\omega\omega_j \ell_j}\label{eq:Amplitudes3}.}
Various sources of loss have been identified (see, for instance \cite{MiklosEtAl2001}). One observes viscous and thermal dissipation in the fluid body (volume loss) and at the resonator boundary layers (surface loss). There is loss due to acoustic wave scattering at surface obstructions, due to the compliance of the chamber walls and  dissipation at the microphone diaphragm. In particular in the region above $\ze{10}{kHz}$ humidity inside the resonator can strongly increase the damping of sound waves, an effect which might be important in the  context of in vivo measurement of blood sugar. Finally, surface roughness can increase surface loss due to viscosity \cite{MiklosLoerincz1989}. The combined effect of the various loss mechanisms can be calculated by adding the individual loss factors.

In the present work only volume and surface loss due to viscosity and thermal dissipation is considered \cite{Kreuzer1977, BaumannEtAl2007}.  The physical parameters used here and in the following are compiled in Table \ref{tab:GasParameters1}. The compliance of the walls of the sample cell is negligible. For modern microphones, like the one used in the setup discussed here (Knowles SPM0404UD5), dissipation at the diaphragm is of no importance.

\begin{table} \caption{Used gas parameters (\textcolor{black}{air}). The values come from the database of the FE software (COMSOL Multiphysics 4, see \cite{ComsolMP}) and correspond to a temperature of ${20}{\rm \grad C}$ and a static pressure of $\ze{1013}{hPa}$. 
\label{tab:GasParameters1}} \label{tab:1}
\begin{tabular}{ll}
\hline\noalign{\smallskip}
\noalign{\smallskip}\noalign{\smallskip}
			density & $\rho=\ze{1.2044}{kg/m^3}$ \\
			sound velocity & $c=\ze{343.2}{m/s}$ \\
			viscosity  & $\eta=\zze{1.814}{-5}{Pa\,s}$ \\
			coefficient of heat conduction  & $\kappa=\zze{2.58}{-2}{W/m\,K}$ \\
			specific heat capacity at constant volume & $c_{\rm V}=\zze{7.1816}{2}{J/kg\,K}$ \\
			specific heat capacity at constant pressure & $c_{\rm p}=\zze{1.0054}{3}{J/kg\,K}$ \\
\noalign{\smallskip}\hline
\end{tabular}
\end{table}

Originally, the differential equation and the surface integrals appearing during the calculation of the loss factors are calculated analytically \cite{Kreuzer1977}. Obviously, only simple geometries can be treated. For realistic profiles of the intensity ${\cal I}(\vec{r})$ the integral (\ref{eq:Amplitudes2}) cannot be calculated without simplifications.  In this paper these quantities are calculated by a FE software and arbitrary geometries and realistic intensities can be used.

Using the method excellent agreement between numerical and experimental results of a T shaped test cell has been obtained \cite{BaumannEtAl2007}. In addition, a shape optimization on the basis of the FE model has been performed which lead to the proposal of an hour glass shaped resonator with an improved signal in comparison to a conventional cell \cite{KostEtAl2011,WolffEtAl2012}.

\subsection{Open resonator}\label{Subsec:OpenResonator}

In this section the differences of an open resonator to a closed resonator are described. The theory comprises advanced mathematics, but we are interested in some basic results only \cite{LevineSchwinger1948}. Levine and Schwinger considered the leaking of acoustic energy at the opening of cylindrical ducts. They assumed the sound wave is generated by a harmonically driven piston inside the pipe \cite{MorseIngard1968}. This system bears a certain similarity with our open resonator (Figure \ref{fig:FrankfurtResonator}). If we ignore the resonance cylinder and regard the sound generation at the skin as similar to the sound generation at the piston, this becomes obvious. The results obtained in \cite{LevineSchwinger1948} can then be used to get an idea what is happening at the open end of the PA resonator.

The $i$-th longitudinal eigenfrequencies of a pipe (radius $R$, length $L$) with one open and one closed end is given by
\nfm{f_i=\frac{(2i-1)c}{4L},\label{eq:eigenfreqOpenPipe1}}
where $i=1, 2, 3, ...$.
The investigations of Levine and Schwinger revealed that $L$ has to be modificated by the end correction $\Delta L=aR$ with $a$ either equal to $0.61$ (unflanged case, meaning the walls of the pipe are of negligible thickness) or $a=0.84$ (flanged case, meaning the walls of the pipe are of infinite thickness). The acoustic pressure drops to zero at a distance $\Delta L$ outside of the opening and the eigenfrequencies are accordingly lower as predicted by Equation (\ref{eq:eigenfreqOpenPipe1}). Figure \ref{fig:FrankfurtResonator} shows a PA cell which approximately represents the flanged case.

The loss due to the leaking of sound energy from the laser beam opening can be described by an additional loss factor \cite{MiklosLoerincz1989}:
\nfm{\ell_j^{\rm rad}=\frac{1}{f_j}\frac{W_j^{\rm rad}}{E_j}.}
$E_j$ is the acoustic energy content of the resonator when the $j$-th mode is excited and $W_j^{\rm rad}$ is defined by a surface integral over the opening:
\nfm{W_j^{\rm rad}=\operatorname{Re} (Y^{\rm op})\int_{S_{\rm op}}{}{\left|p_j\right|^2}{{\;\rm d}S,}}
$\operatorname{Re} (Y^{\rm op})$ denotes the real part of the specific admittance of the opening.

In \cite{LevineSchwinger1948} the low frequency behavior of the acoustic impedance (the inverse of the admittance) of the opening has been derived:
\nfm{Z^{\rm op}=\frac{\rho c}{S_{\rm op}}kR\left({\rm i}a_1+a_2 kR\right).}
The wave number $k$ is complex since the waves suffer attenuation. The numbers $a_1$ and $a_2$ depend on the frequency. In practice one often uses $a_1\approx 0.6$ and $a_2\approx 0.25$ as an approximation \cite{MiklosEtAl2001}.

In leading order of $kR$, i.e. in the low frequency limit, the impedance is purely imaginary. In this limit the sound wave is completely reflected near the opening and suffers a phase shift of $\pi$. When the frequency is increased sound energy begins to leak out of the opening.

It seems that we could use the  above equation for $Z^{\rm op}$ to obtain the admittance of the opening and to calculate the radiation loss factor. Unfortunately, the theory is applicable for  $\omega < \omega_{\rm crit}=2\pi f_{\rm crit}$ only. For the resonator considered in this paper the critical frequency is about $f_{\rm crit}=\ze{25}{kHz}$ and much lower than the frequency used in experiment. Above $f_{\rm crit}$ radial and azimuthal modes contribute in the case of a cylindrical geometry \cite{Weinstein1974}.

The general tendency is that the amount of energy that leaks out of the opening of the resonator increases with frequency. However, it not possible to state how much acoustic power is lost in comparison to the other loss mechanisms discussed earlier. In \cite{MiklosLoerincz1989} it has been pointed out that openings not necessarily increase the loss substantially.


\section{Results}\label{Results}

In \cite{PleitezEtAl2013} the response function of the open resonator of Figure \ref{fig:FrankfurtResonator} has been determined experimentally. In addition, Pleitez and coworkers have measured the response of the cell after sealing the opening with a window. In both cases a broad band absorbing material (glassy carbon black) has been used as a reference sample. The laser beam is focused to a point slightly off the sample so that the sample is irradiated across the whole skin opening.

We present results for the same cells obtained from a FE model and compare them to the experimental response functions\footnote{\textcolor{black}{The experimental data used in this paper are not identical to those used in \cite{PleitezEtAl2013}. The deviation is minor.}}. As explained in the previous section we are not able to determine the loss factor due to sound radiation. Therefore, a comparison of the measured and the calculated response should indicate how much radiation loss in the open cell contributes.

\subsection{Closed resonator}\label{Subsec:ResultsClosedResonator}

The FE model is well tested for the case of a closed resonator \cite{BaumannEtAl2007} and, therefore, it is reasonable to begin here. All boundaries of the closed cell are assumed to be sound hard.

As mentioned in Section \ref{Subsec:ClosedResonator} the source term for the excitation of sound vanishes everywhere outside a small region near the surface representing the skin of the proband or the surface of the carbon black sample, respectively. Instead it has been used ${\cal I}(\vec{r}) = \mbox{const}$ in the small cylinder adjoining this surface. This is for technical reasons and has no significant influence on the results, as has been checked by varying the shape and the size of the region where the heat source is active.

The microphone is mounted flush at the end of the resonance cylinder and the diaphragm represents its ending.  Since the exact location of the diaphragm is not known, we have conducted a parameter study, which revealed that an effective length of the resonance cylinder of $\ze{7.4}{mm}$ gives the best match with experimental results.

\begin{figure}
\includegraphics[width=0.95\textwidth,trim=80 20 70 20,clip]{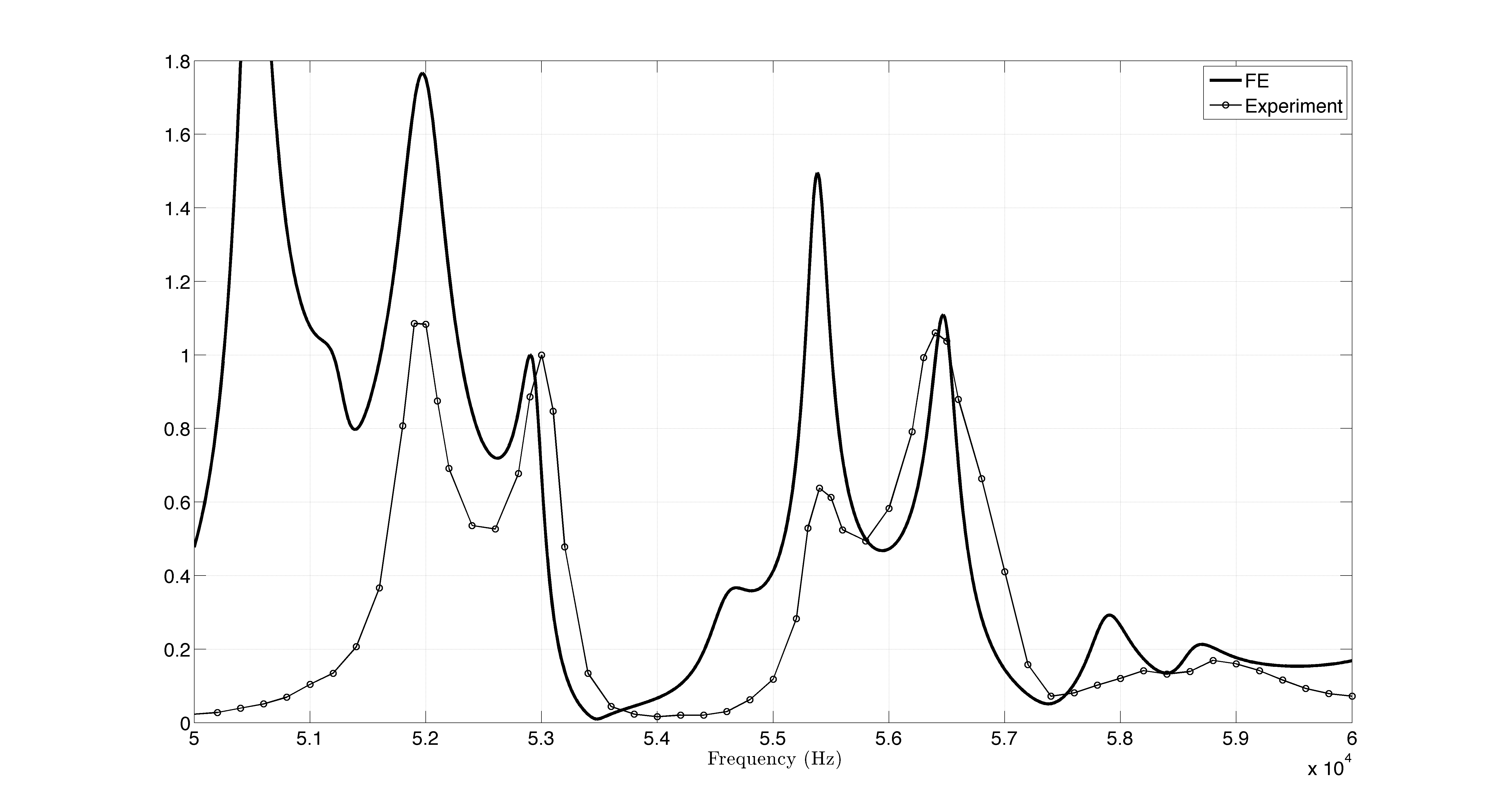} \caption{Response function of closed resonator. The thin line with circles show the experimental result, the thicker full line the FE result. The response axis is in arbitrary units. The curves have been rescaled such that the resonance amplitudes for the resonance at about $\ze{53}{kHz}$ coincide.} \label{fig:ARFclosedLR7k4mm}
\end{figure}

\begin{figure}
\includegraphics[width=0.15\textwidth,trim=50 30 60 20,clip]{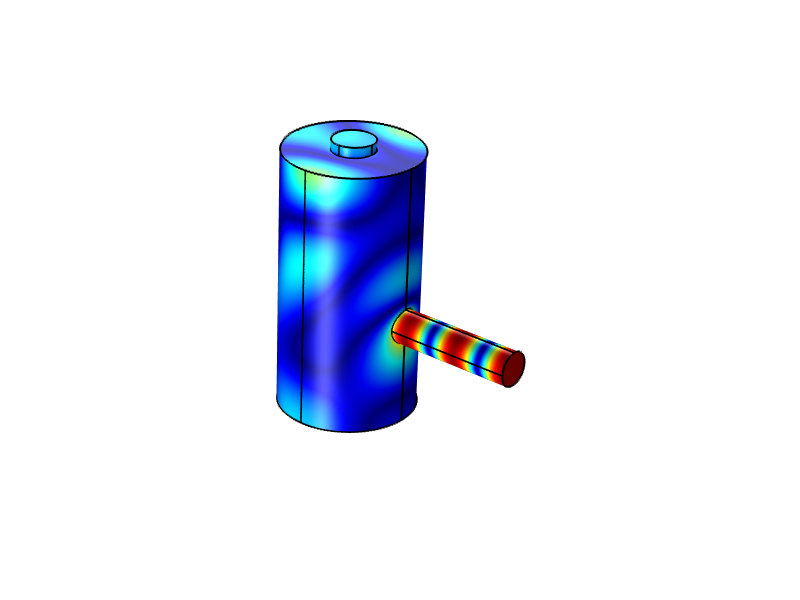}
\includegraphics[width=0.15\textwidth,trim=50 30 60 20,clip]{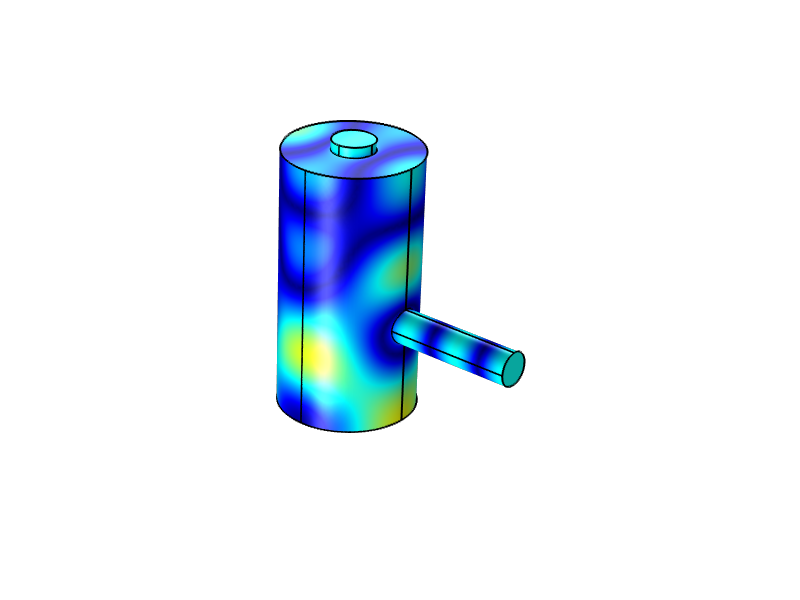}
\includegraphics[width=0.15\textwidth,trim=50 30 60 20,clip]{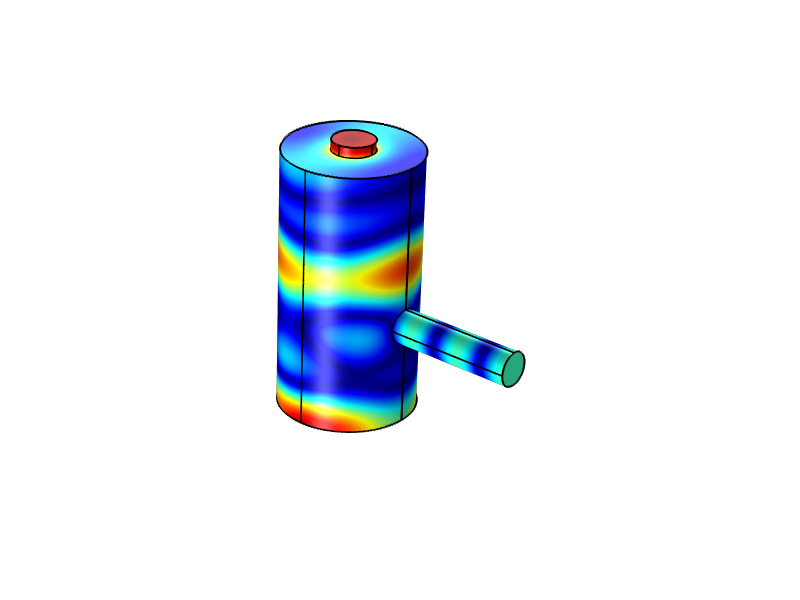}
\includegraphics[width=0.15\textwidth,trim=50 30 60 20,clip]{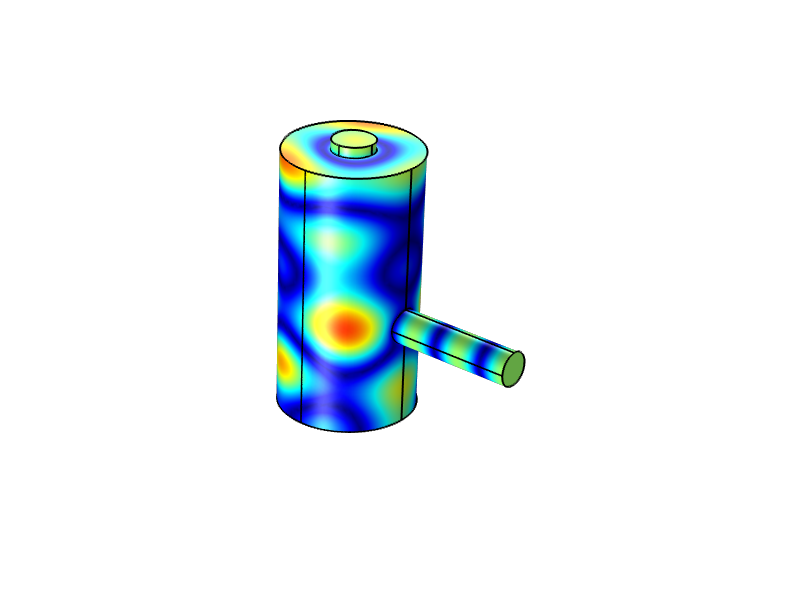}
\caption{Dominant modes of the closed resonator. Depicted is $|p|$. Dark blue corresponds to $|p|=0$, red to the maximal value of $|p|$. The absolute values of $|p|$ are of no significance. The corresponding frequencies are $\ze{52.0}{kHz}$, $\ze{52.9}{kHz}$, $\ze{55.4}{kHz}$ and $\ze{56.5}{kHz}$.}
 \label{fig:DominanteModesClosedResonator1}
\end{figure}

In Figure \ref{fig:ARFclosedLR7k4mm} the experimental response function of the closed resonator is depicted together with the result of the simulation. It is obvious that the main resonances observed in experiment also appear in the simulation. The resonance frequencies are predicted with very good precision (deviation $<0.2\%$). The corresponding modes are displayed in Figure \ref{fig:DominanteModesClosedResonator1}. For at least some of the resonances the width observed in experiment is larger than the width of the calculated resonances. This indicates loss mechanisms which are not included in the model (see section \ref{Subsec:ClosedResonator}).

One can speculate on the influence of surface roughness of the cell walls. As mentioned previously the roughness of the cell walls can increase surface loss due to viscosity. In cylinder cells with a length to diameter ratio significantly larger than one, the roughness of the surface is important for longitudinal modes and less important for radial modes. This is due to the fact that in longitudinal modes the velocity of the fluid particles is along the long cylinder barrel while in radial modes it is along the small basal and top area of the cylinder. In T shaped cells the modes cannot be classified into these categories. Still, modes can have a more longitudinal or a more radial character. The first and the third mode of Figure \ref{fig:DominanteModesClosedResonator1} correspond to weaker resonances in experiment than one would expect from the calculated response. In both modes one can identify a longitudinal component: For the first mode in the resonance cylinder and for the third mode in the main cylinder. This reasoning might serve as an explanation for the difference in calculated and measured response functions. 

\begin{figure}
\includegraphics[width=0.15\textwidth,trim=50 30 60 20,clip]{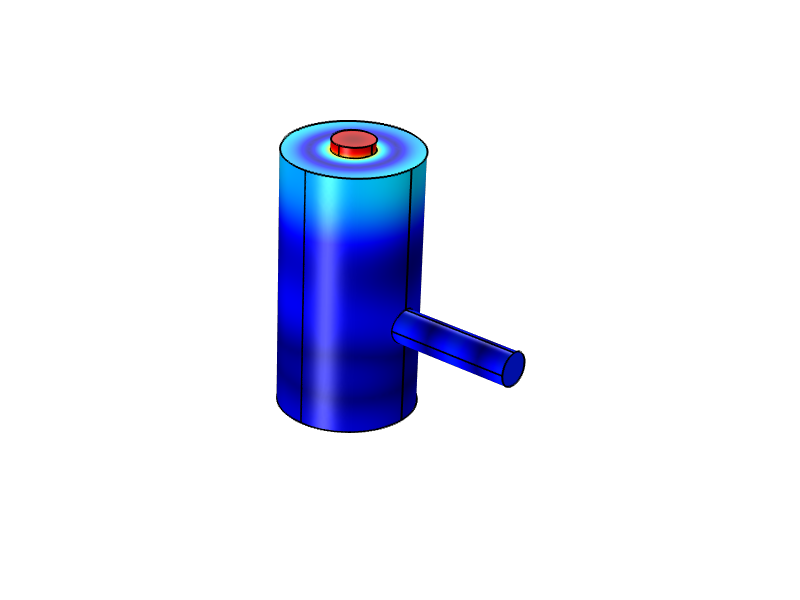}
\includegraphics[width=0.15\textwidth,trim=50 30 60 20,clip]{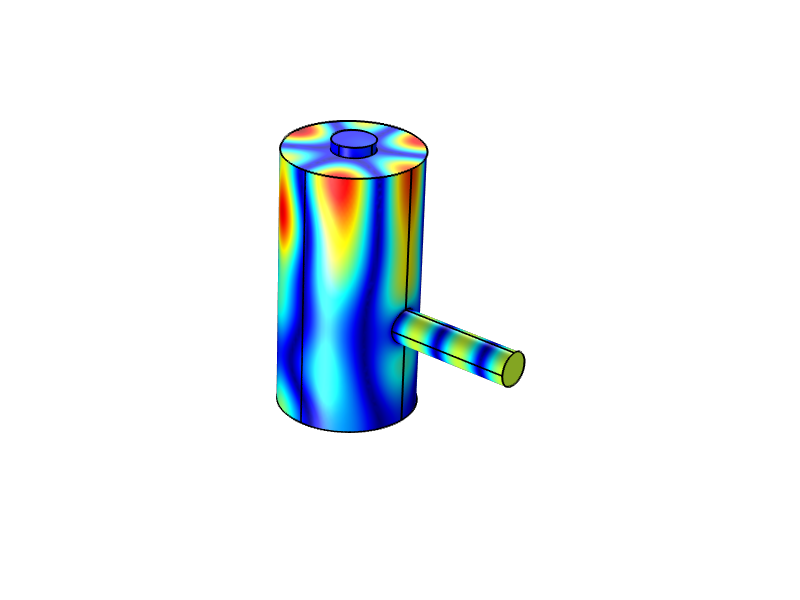}
\caption{Modes corresponding to resonances which do not appear in in the experimental response curve of the closed resonator. Depicted is $|p|$. The frequencies are $\ze{50.6}{kHz}$  and $\ze{57.9}{kHz}$. The second mode clearly has the character of an azimuthal mode of the main cylinder.}
 \label{fig:AdditionalModesClosedResonator}
\end{figure}

\begin{figure}
\includegraphics[width=0.75\textwidth,trim=0 0 0 0,clip]{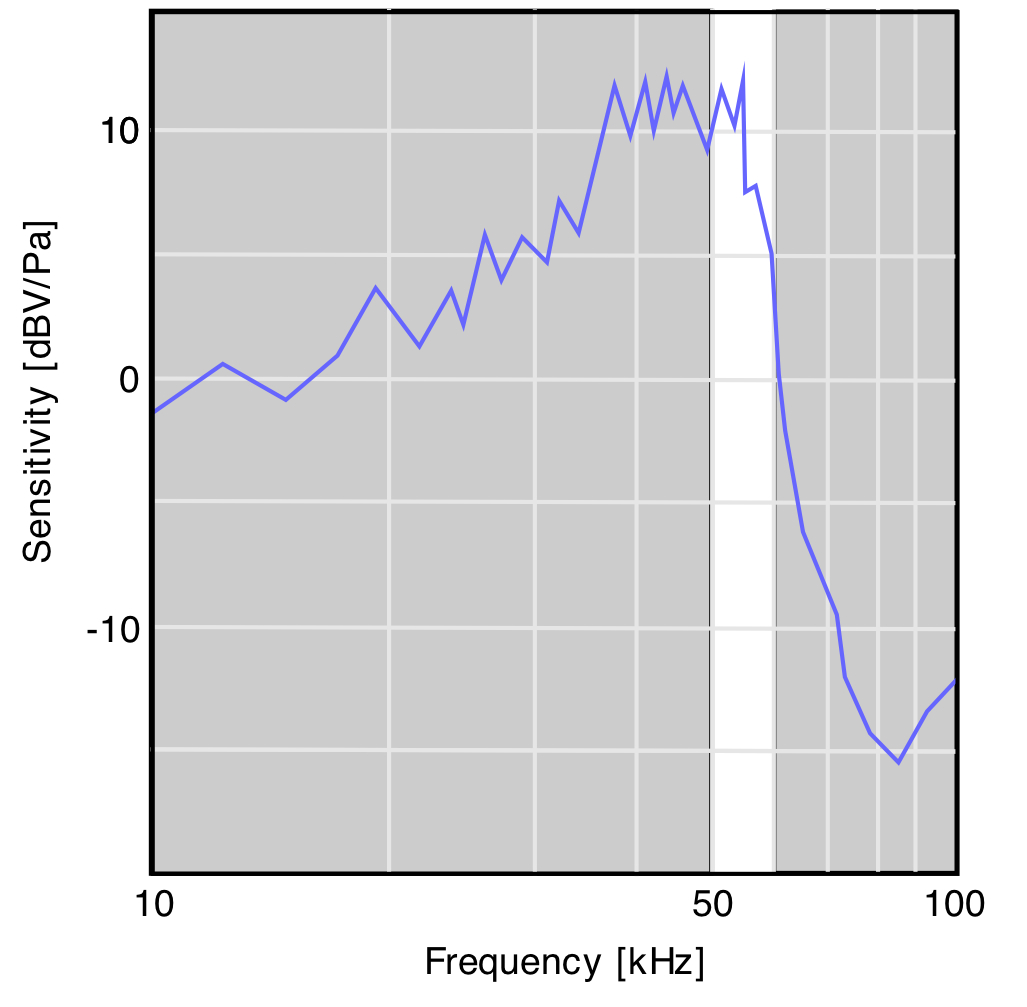} \caption{Sketch of the response curve of the microphone (Knowles Acoustics SPM0404UD5). Apparently there is a substantial drop of the curve in the relevant frequency range from $50$ and $\ze{60}{kHz}$.} \label{fig:SPM0404UD5}
\end{figure}

The FE model predicts a strong resonance around $\ze{50.6}{kHz}$ and a small resonance at about $\ze{57.9}{kHz}$ which are not present in the measured response function (Figure \ref{fig:AdditionalModesClosedResonator}). The absence of the latter can possibly be explained by a steep drop of the microphone sensitivity of about $\ze{20}{dB}$ in the frequency range $\ze{55}{kHz}<f<\ze{60}{kHz}$ (see Figure \ref{fig:SPM0404UD5}). We are not able to offer an explanation for the missing resonance at $\ze{50.6}{kHz}$. Most likely some details of the cell geometry have not been encompassed in the FE model and this difference is responsible for the suppression of the resonance at the microphone.

\subsection{Open resonator}\label{Subsec:ResultsOpenResonator}

Since we are not able to include an impedance boundary condition that describes radiation loss in our model we chose to model the opening in the simplest possible way by a sound soft boundary condition. All other boundary conditions and the source term are identical to section \ref{Subsec:ResultsClosedResonator}. The open boundary is not included in the surface integrals, which are used to calculate the surface loss factors \cite{BaumannEtAl2007}.

\begin{figure}
\includegraphics[width=0.95\textwidth,trim=80 20 70 20,clip]{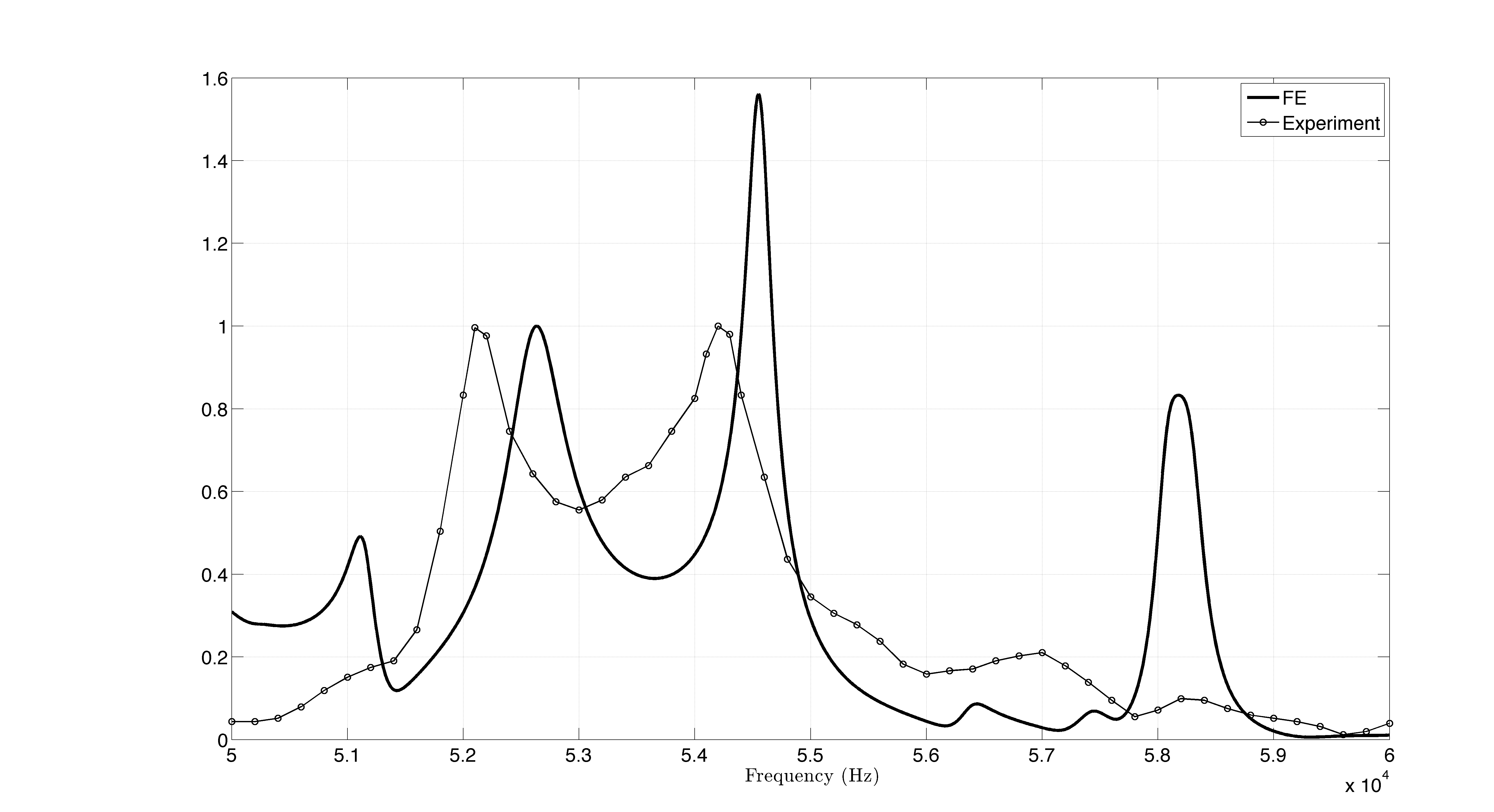} \caption{Response function of open resonator.  The response axis is in arbitrary units. The curves have been rescaled such that the resonance amplitudes for the  resonance at about $\ze{52}{kHz}$ coincide.} \label{fig:ARFopenLR7k4mm}
\end{figure}

\begin{figure}
\includegraphics[width=0.15\textwidth,trim=50 30 60 20,clip]{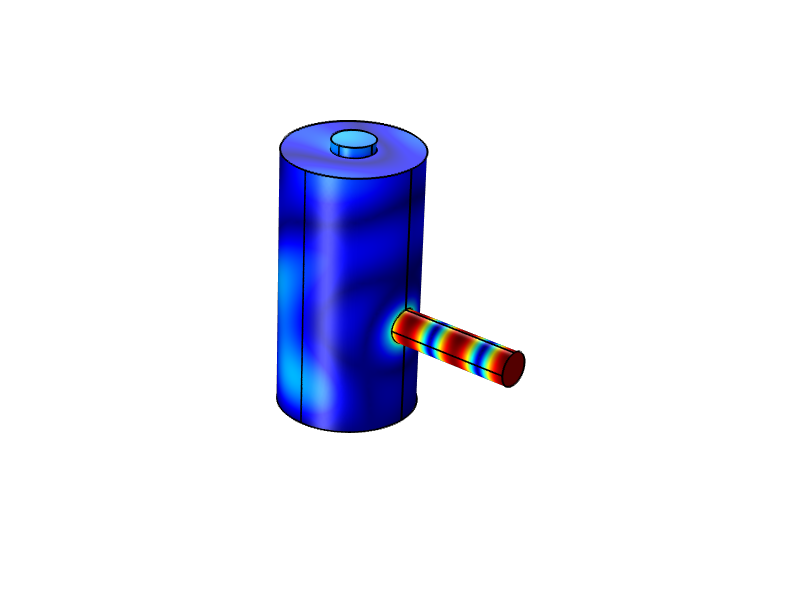}
\includegraphics[width=0.15\textwidth,trim=50 30 60 20,clip]{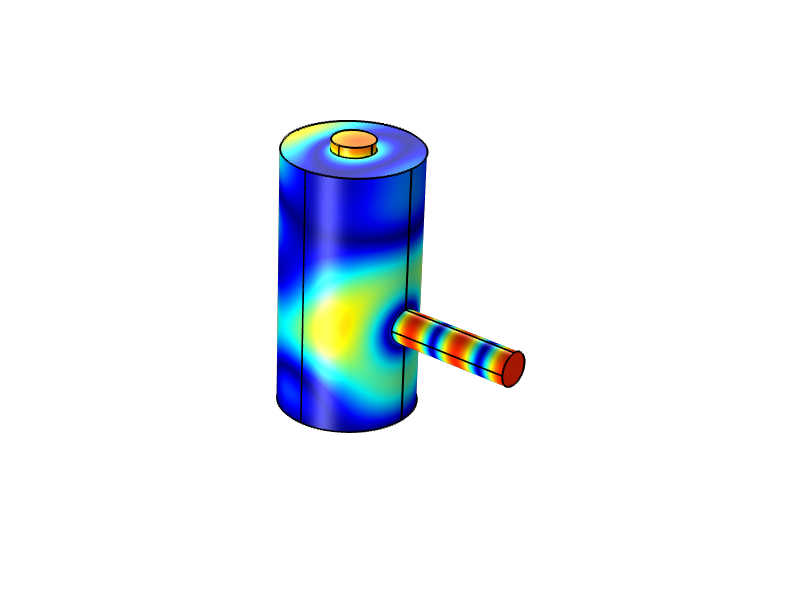}
\caption{Dominant modes of the open resonator. Depicted is $|p|$. The frequencies are $\ze{52.7}{kHz}$ and $\ze{54.6}{kHz}$.}
 \label{fig:DominanteModesOpenResonator1}
\end{figure}

In Figure \ref{fig:ARFopenLR7k4mm} the experimental and the calculated response function are depicted. Starting from the high frequency end there is a pronounced resonance in the simulation curve at about $\ze{58.2}{kHz}$ which is not present in the experimental curve. Again, this might be attributed to the microphone sensitivity as in the case of the closed resonator. At  $52.7$ and $\ze{54.6}{kHz}$ one observes strong resonances in the simulation result. The corresponding modes are displayed in Figure \ref{fig:DominanteModesOpenResonator1}. The experimental response function shows two similar resonances at slightly lower frequencies (deviation about $\ze{1}{\%}$). This is not surprising since at the open end one has to expect an end correction which leads to larger wavelengths, i.e. lower frequencies.

Without supporting the statement quantitatively it is apparent that the deviation of height and width between calculated and measured resonance amplitudes is similar to the one of the closed cell. If the loss of acoustic energy due to radiation at the laser beam opening of the resonator would be substantial, the accordance of the measured and calculated response should be poorer in the case of a open cell. Finally, the calculated response function exhibits a resonance at about $\ze{51.1}{kHz}$, which has no counterpart in the experimental result. The discrepancy might result from the same reason, which has been offered in Section \ref{Subsec:ResultsClosedResonator} for the closed resonator.


\section{Conclusion}\label{Conclusion}

We applied a FE model originally designed for the calculation of the microphone signal of PA sensors with closed resonators for the determination of the response function of an open resonator. At the opening a simple sound soft boundary condition has been assumed. Comparison of numerical and experimental results reveal a fairly good concordance. The fact that some measured resonances are spectrally wider than the calculated ones shows that not all relevant loss mechanisms are included in the model. However, the deviation of resonance qualities is similar for closed and open resonator. This is an indication that loss due to sound radiation at the open boundary is not a dominant loss effect. In order to get a good prediction, it is important to know the geometrical data with high accuracy. It might also be important to consider the roughness of the cell walls. Further studies should take all details of the resonator shape into account.

\vspace{3mm}
\noindent {\bfseries Acknowledgements:} This research was supported by the Free and Hanseatic City of Hamburg and the  Hamburg University of Applied Sciences. We are indebted to Prof. W. M\"antele and his team at the Institute of Biophysics, Goethe University Frankfurt, for supplying design and experimental data. Furthermore, we like to thank Dr. Lars Duggen from the University of Southern Denmark for discussions and support.

\bibliographystyle{unsrt}

\end{document}